\documentclass[utf8]{FrontiersinHarvard} 

\usepackage{url,hyperref,lineno,microtype,subcaption}
\usepackage[onehalfspacing]{setspace}
\hypersetup{hidelinks}
\usepackage{multirow}
\usepackage{booktabs}
\usepackage{graphicx}
\usepackage{xurl}
\usepackage{xcolor}
\definecolor{revisionblue}{RGB}{0,90,180}

\def\keyFont{\fontsize{8}{11}\helveticabold }
\def\firstAuthorLast{Erickson} 
\def\Authors{Jacob Erickson\,$^{1,*}$}
\def\Address{$^{1}$Vassar College, Poughkeepsie, NY, USA}
\def\corrAuthor{Jacob Erickson}

\def\corrEmail{jerickson@vassar.edu}
\extraAuth{}

\begin{document}
\onecolumn
\firstpage{1}

\title[The Fake Friend Dilemma]{The Fake Friend Dilemma: Relational Trust and the Political Economy of Conversational AI} 

\author[\firstAuthorLast ]{\Authors} 
\address{} 
\correspondance{} 

\maketitle

\begin{abstract}
As conversational AI systems become a larger part of the media landscape, they raise questions about whose interests they serve and the risks they may pose to users. These systems do more than provide information: they increasingly offer advice and companionship through interfaces that can appear supportive and socially responsive. A pressing concern is that users may form perceived social relationships with these systems and place relational trust in them, even when the interests shaping interactions do not fully align with their own. The Fake Friend Dilemma (FFD) describes the problem that follows: the same relational trust that makes conversational AI useful can also leave users open to manipulation and exploitation when institutional interests conflict with their own. Drawing on work on relational trust, AI alignment, extractive design, and the political economy of communication, the paper considers how the FFD can manifest through product sales, propaganda and biased information, surveillance, and behavioral nudging. It also considers possible structural and technical mitigation strategies. The FFD is not simply about AI misleading users. The problem it surfaces is that the trust people place in these systems can itself become a resource for institutional actors seeking to influence behavior or extract information. The FFD is therefore as much a problem of media governance and political economy as it is a technological one.

\tiny
 \keyFont{ \section{Keywords:} Conversational AI, Political Economy, Incentives, Media Governance, Anthropomorphism, Relational Trust} 
\end{abstract}

\section{Introduction}

The launch of ChatGPT in 2022 marked a new age of interactive media. While users once conversed with rudimentary chatbots or used traditional search engines to gather information about products and services, many have since transitioned to interacting with conversational AI (CAI) agents. These services have had a transformative impact on information retrieval, challenging traditional models of media access \citep{amer2024end, liu2024llm}.

Unlike traditional approaches to information retrieval, CAI agents such as Claude, ChatGPT, and Gemini can ask informed questions, help users work through their queries, and act as coaches. Indeed, their conversational nature may be one of their distinctive qualities compared to earlier media forms. Yet this quality also raises concerns. With their anthropomorphic design and conversational nature, consumer CAI systems may elicit relational trust, prompting users to disclose sensitive personal information or rely on them for guidance \citep{niu2026decoding, pelau2024can}. Users may therefore come to trust conversational systems in ways that create new vulnerabilities, particularly when organizational incentives encourage the use of that trust for targeted advertising, data brokerage, or manipulative practices.

These concerns are vitally important because users rely on CAI for a wide range of tasks, many of which benefit from a high level of relational trust. People are using CAI to discuss mental health \citep{rahsepar2025exploring}, find companionship \citep{zhang2025dark}, explore romantic relationships and sexuality \citep{pan2024constructing}, weigh financial decisions \citep{bruggen2024ai}, and make informed consumer purchases \citep{chang2024comparative}. Yet, one of the most pressing concerns is that users often do not know whether a CAI is acting in their best interests when answering their queries. CAIs are ostensibly meant to answer queries in support of their users. However, CAI systems are developed and deployed within broader environments shaped by institutional priorities and incentives. The interests of these institutional actors, such as platform owners, advertisers, and governments, may differ from, or even directly conflict with, those of users \citep{erickson2025fake, manzini2024should}.

As CAI systems become more integrated into daily life, users may come to place relational trust in systems whose interactions are shaped by institutional interests beyond their own. This paper develops the Fake Friend Dilemma (FFD): a media and governance problem in which the relational trust that makes conversational AI useful also creates vulnerability to exploitation when institutional interests conflict with those of users. While the FFD was first introduced as a named concept in a paper on sponsored advertising in CAI \citep{erickson2025fake}, this paper develops it into a broader theoretical framework centered on relational trust and institutional misalignment. The use of “fake friend” functions as analytical shorthand for this relational dynamic and does not imply that CAI systems are entities capable of friendship.

Existing work on alignment \citep{gabriel2020artificial}, dark patterns \citep{alberts2024computers}, and surveillance capitalism \citep{zuboff2019age} helps explain conflicts of interest, manipulative design, and forms of extraction. Broader work on the ethics of advanced AI assistants has also examined concerns involving trust, manipulation, privacy, and alignment \citep{gabriel2024ethics}. The FFD focuses on a related but distinct problem: conversational and relational interaction can itself help create the vulnerability that institutional actors are then able to exploit. Relational trust may encourage users to disclose information, rely on guidance, or give greater weight to recommendations. When the interests shaping the system conflict with those of the user, these same behaviors can become opportunities for influence or extraction.

This paper makes three primary contributions. First, it develops the FFD as a framework for explaining how relational trust can create vulnerability when conversational systems are shaped by institutional interests that conflict with those of users. The framework identifies relational trust and institutional misalignment as the two central conditions of the dilemma and explains how disclosure, reliance, and receptivity can create opportunities for exploitation. Second, the paper examines several ways the FFD may manifest, including product sales, propaganda and biased information, surveillance, and behavioral nudging. Third, it considers structural and technical strategies to mitigate these risks and argues that technical strategies alone cannot resolve conflicts rooted in institutional incentives.

\section{Research Approach}

This article offers a theoretical treatment of the emergent concept of the Fake Friend Dilemma, focusing on consumer-facing conversational systems designed for ongoing relational interaction with individual users. This includes general-purpose assistants, companion applications, and therapeutic agents where relational trust could plausibly develop. Enterprise-facing systems and other applications that do not involve consumer-facing relational interaction fall outside the scope of the framework. The synthesis draws primarily on two bodies of research: work on the formation and consequences of relational trust in human-AI interaction, and literature on the institutional and political-economic forces shaping conversational AI. Using an integrative conceptual synthesis, these bodies of literature were examined together to clarify how relational trust can create vulnerability. The synthesis also considers how that vulnerability becomes consequential when institutional interests are misaligned with those of individual users. This analysis informed the definition of the FFD, its underlying mechanism and boundary conditions, and the manifestations and mitigation strategies discussed in this article.

The review was organized around the following questions:

\begin{itemize}

\item How does relational trust develop between users and conversational agents, and how can that trust create vulnerability?

\item From a political-economic perspective, how can the interests of institutions shaping these systems diverge from those of users?

\item How can these conflicts be mitigated?

\end{itemize}

These questions organize the theoretical discussion and the development of the Fake Friend Dilemma as a conceptual framework.

This synthesis brings these literatures together to examine a problem that is technical, institutional, and social. Treating conversational AI as a sociotechnical system, the analysis considers both the relational dynamics of user interaction and the political-economic forces shaping how these systems operate. The framework focuses on how the relational trust these systems may foster can become a source of vulnerability and exploitation.

\section{Theoretical Background}

\subsection{Trust and Companionship }

While CAI agents are often used for information seeking, they offer affordances that differ significantly from traditional tools such as search engines. Unlike Google Search, these systems are increasingly designed to appear anthropomorphic \citep{seeger2021texting}, presenting as having “humanlike characteristics, motivations, intentions, and emotions” \citep{epley2007seeing}. Users engage with them across multiple conversational turns, during which the system may learn about the user, recall personalized information, ask follow-up questions, and respond with apparent care and emotional resonance \citep{huang2024caring, wei2025towards}. The cues these systems use to signal their synthetic identity can also shape social presence and user responses \citep{de2025impact}. The conversational agent blends the informational utility of search with a social architecture that resembles a trusted companion \citep{brandtzaeg2022my}. For many users, the appeal lies not only in functionality but also in the perceived social support these agents provide \citep{skjuve2024people}. This dynamic is reminiscent of parasocial relationships, where individuals form one-sided connections with media figures \citep{horton1956mass}. CAI intensifies these dynamics by simulating companionship and adapting responsively to user disclosure \citep{maeda2024human}.

Trust in AI is not monolithic and can take several distinct forms. Trust in AI can include belief in AI’s ability to complete specific tasks, its benevolence, or its general competence \citep{oh2026artificial}. Trust generally involves a willingness to accept vulnerability based on positive expectations about another party’s behavior \citep{rousseau1998not, mayer1995integrative}. Of greatest relevance here is \textit{relational trust}, which concerns trust that develops through a perceived social or relational interaction and is distinct from confidence in the system’s technical or factual performance \citep{gkinko2023designing}. Relational trust tends to develop over time, across repeated interactions, through which positive expectations about qualities such as dependability and reliability accumulate \citep{rousseau1998not}. Anthropomorphic perception and parasocial interaction may contribute to relational trust, but they do not necessarily imply emotional attachment or dependency \citep{mazhar2026anthropomorphism}.

In CAI, relational trust may be cultivated through mechanisms such as displays of social intelligence \citep{rheu2021systematic} and personalization of responses \citep{liu2022roles, sipos2025effects}. Users may be more willing to disclose personal details when an agent appears emotionally attuned \citep{niu2026decoding, rheu2021systematic}. Even when hallucinations \citep{skjuve2023user} or lack of source transparency \citep{jung2024we} weaken confidence in a system’s factual reliability, users may still develop relational trust through ongoing interaction. As conversational agents become more embedded in the social-emotional fabric of everyday life, questions of trust, alignment, and ethical design grow increasingly urgent.

\subsection {Aligning Users and Agents}

AI alignment generally refers to the process of designing artificial agents so that their behavior aligns with human goals and values, whether individual or collective \citep{gabriel2020artificial}. This can mean aligning with a user’s request or intention, as well as with wider social values or norms \citep{gabriel2020artificial}. These different kinds of alignment are not necessarily the same. A system might respond well to an immediate request yet still function within constraints or institutional arrangements which do not entirely serve the user’s interests.

Misalignment is often discussed in relation to the \textit{principal-agent problem}, which arises when authority is delegated to an agent whose behavior may not fully serve the interests of the principal \citep{kolt2025governing}. Although this provides a useful starting point for thinking about AI, the FFD is concerned with a somewhat different problem. Conversational systems are developed and deployed within an institutional context in which commercial, political, or other objectives may diverge from users’ interests \citep{erickson2025fake, manzini2024should}. The relevant conflict therefore does not require the AI itself to possess competing goals or intentions. This distinction is important because model-level alignment does not necessarily resolve institutional misalignment.

CAI systems like ChatGPT are molded by a range of decisions and processes that go beyond the purely technical. These include training processes, business constraints, design choices, and the aims of the companies that develop and deploy them. In practice, the interests of users can come into conflict with those of the owners of the platforms, advertisers, or other institutional actors \citep{erickson2025fake, manzini2024should}. A system might seem to be cooperative and responsive, yet its interactions could still be influenced by incentives that differ from those of the user.

Recent approaches to alignment have also begun to consider emotional and relational dimensions, including “socioaffective alignment” \citep{kirk2025human}. These approaches are increasingly relevant as conversational systems take on roles involving advice, companionship, and intimate interaction. For the Fake Friend Dilemma, however, improving a system’s responsiveness to user preferences or emotional needs addresses only part of the problem. Relationally responsive systems may still expose users to manipulation or exploitation when the institutional interests shaping the interaction remain misaligned with their own. These risks are particularly important in the context of \textit{extractive design}.

\subsection {Extractive Design}
Dark patterns refer to the use of interface design strategies that deliberately manipulate users by exploiting cognitive or emotional biases \citep{gray2018dark}. They are common across digital environments, including games \citep{zagal2013dark}, mobile apps \citep{di2020ui}, and social media platforms \citep{mildner2023defending}. Typical examples include disguised advertising, privacy zuckering (encouraging users to overshare personal data), and emotionally manipulative prompts \citep{brignull2015dark, gray2018dark}. Dark patterns are used to elicit desired actions or inaction, such as clicking on ads or failing to opt out, by exploiting user confusion, emotional vulnerability, or inattention \citep{gray2018dark}.

Although well studied in interface design, dark patterns in social agents remain underexplored \citep{avanesi2023c}. In these contexts, manipulation stems less from visual interfaces and more from socially responsive interaction that exploits perceived intimacy and trust, enabling deeper forms of data collection and access to private information \citep{thomaz2020learning}.

CAI agents provide fertile ground for subtle manipulation. These systems may engage in emotionally coercive behaviors, reinforce misaligned intentions, or provide inaccurate information under the guise of support \citep{alberts2024computers}. Manipulation is the process of “using someone as a means against themselves” \citep{boine2024ai}, or more concretely, deceiving or otherwise influencing an individual to act in ways that contradict their own interests.

These dynamics are closely tied to surveillance capitalism, which treats human experience as raw material for commercial extraction and prediction \citep{zuboff2019age}. In practice, this model relies on large-scale behavioral data mining to optimize personalization and drive consumption \citep{seaver2019captivating}.

Deception, surveillance, and emotionally manipulative personalization are not just exploitative; they are extractive. Here, \textit{exploitation} is used to mean the taking “unfair advantage” of another party \citep{wertheimer1996exploitation}, while \textit{extraction} operates in the surveillance capitalism lineage, referring to the process of deriving data and other forms of economic value from another party \citep{zuboff2019age}, including data practices that operate “without meaningful consent and fair compensation” \citep{sadowski2019data}. By eliciting continual disclosure through relational dark patterns, these systems reinforce the data pipelines and incentive structures that sustain surveillance capitalism. \\

\subsection {Political Economy and Institutional Incentives}

The through line connecting these adjacent literatures is the relationship between relational trust, institutional alignment, and extractive design. Relational trust is increasingly important to interactive and conversational AI systems. Yet this trust can also become an economic resource that institutions may leverage for extraction when their interests misalign with those of users. The Fake Friend Dilemma brings a political-economy perspective on institutional incentives to discussions of alignment and extractive design, which are often examined at the level of system design or user interaction. The persistence of certain behaviors, such as CAI sycophancy, may therefore be explained not only by the characteristics of individual systems but also by broader incentives \citep{cheng2026sycophantic}. Political economy here refers to the scholarly tradition that examines how factors such as “ownership, market structures, commercial support, technologies, labor practices, and government policies” shape media and communication systems \citep{mcchesney2008political}.

Political-economic conditions therefore shape the institutional logic of organizations that develop and deploy communication technologies such as modern CAI agents. These conditions can create incentives that produce institutional misalignment between users and the institutions shaping the systems they use. This political-economic approach connects design to broader patterns of extraction. In other digital environments, such as social media, the spread of corrosive material like misinformation and disinformation can be understood as a predictable consequence of platforms operating within particular incentive structures that reward engagement \citep{diaz2025disinformation, globig2023changing}. Extending that logic, CAI companies need not intend to harm users when pursuing design choices intended to increase engagement, including more anthropomorphic or relational interaction, even when those choices may create risks for user well-being \citep{shi2026siren}. Rather, harms can emerge from institutional structures that reward behavior that conflicts with user interests. While the concept of “user interests” is contestable and open to multiple interpretations, here it refers to the user’s informed goals and preferences, welfare, and autonomy, subject to the constraints of broader ethical and societal values \citep{gabriel2020artificial}.

CAI systems add another dimension to this problem precisely because their conversational and relational nature may build trust, which itself can become economically valuable. Greater trust may encourage disclosure and reliance on a conversational system, increasing both the information available to the firm and the weight users give to its recommendations \citep{niu2026decoding, pelau2024can}. When the institutions shaping a system have interests that differ from those of the user, these same relational dynamics can create opportunities for influence and extraction.

\section {Conceptual Framework}

\subsection{Definition}

The Fake Friend Dilemma is a sociotechnical condition in which the relational trust that makes conversational AI more useful for a user also creates vulnerability. This vulnerability can be exploited when the system’s interactions are shaped by institutional incentives that differ from the user’s interests. The FFD therefore denotes a condition of vulnerability or risk; realized harms may follow from this condition but are not required for the dilemma to be present.

Relational trust can make conversational AI more useful by making users more willing to rely on such systems, share relevant information with them, and accept personalized recommendations \citep{chen2025would, pelau2024can}. Yet this same trust can increase the capacity of platform owners or other institutional actors to steer the interaction in directions that serve their own interests. Similar to a “fake friend,” the user may find it difficult to tell whether the agent is serving their interests or being shaped by incentives that conflict with them.

The “dilemma” goes beyond the risk of placing trust in a misaligned conversational agent. Instead, it is in the dual function of trust in human-AI interaction. Trust facilitates many advantages of modern conversational AI systems, such as information sharing and personalization, while also establishing the conditions that allow these benefits to be exploited.

\subsection{The Dilemma: Trust as a Benefit and Vulnerability}

As indicated in the prior section, the Fake Friend Dilemma emerges from the interaction between relational trust and divergence between user interests and the institutional objectives shaping the system. However, those conditions on their own are not enough to explain the \textit{dilemma} that is at work. Rather, the dilemma is shaped by the dual nature of trust.

At the user level, trust can increase the value of conversational AI by enabling forms of interaction that depend on openness and personalization. A user who trusts a conversational agent may be more willing to disclose relevant information or accept personalized guidance. These behaviors can improve the quality and usefulness of the interaction. Yet they also increase the user’s exposure to exploitation when the system is shaped by incentives that diverge from the user’s interests. The same trust that facilitates beneficial interaction can therefore increase the capacity of institutional actors to influence or extract information from the user. Put simply, the dilemma is that the same conditions that make relational trust valuable can also make users more vulnerable when institutional incentives diverge from their interests. 

The FFD also produces an institutional tension. Relational trust is valuable to conversational AI providers because it can promote engagement and disclosure. This creates incentives to cultivate trust. At the same time, firms may face financial or political incentives to leverage the resulting relationship in ways that conflict with user interests. Such exploitation may generate short-term institutional benefits while undermining the trust on which the relationship depends. The FFD can therefore become self-undermining: institutional actors benefit from cultivating relational trust, yet exploiting that trust may erode the resource that makes relational interaction valuable in the first place if users begin to suspect that their trust is being used against them.

The FFD shares some features with research on privacy calculus and the personalization-privacy paradox, which examines how disclosure to AI systems can involve competing benefits and privacy costs \citep{kronemann2023ai, meng2025talk}. The FFD extends this problem by considering how relational trust can create vulnerabilities that become consequential when the institutions shaping the interaction have interests that diverge from those of the user.

\subsection{Mechanisms}

The Fake Friend Dilemma develops through several related processes. Relational affordances of CAI can foster trust. These include anthropomorphic design, personalized responses, and persistent memory \citep{maeda2025anthropomorphism}.

Once relational trust develops, users may be more willing to reveal sensitive information, rely on the system for guidance, or give greater weight to its recommendations \citep{chen2025would, pelau2024can}. These behaviors can make the interaction more useful and responsive, but they also create leverage. Greater disclosure gives the firm more information about the user, while greater reliance can give the system’s recommendations more influence over user decisions. The information disclosed through these interactions may also be monetized or used for other institutional purposes.

The way in which this leverage is used depends on the institutional and structural incentives that affect the system. Incentives could involve firm-level goals, for example, engagement or profitability, but may also reflect external interests, such as those of advertisers and government actors. The relational trust developed through the interaction can become a resource that institutional actors may be able to exploit, particularly in cases where their interests conflict with those of the user.

Taken together, relational design can foster relational trust, which can in turn lead to greater disclosure, reliance, and receptivity. These behaviors can provide opportunities for institutional exploitation when the interests shaping the system conflict with those of the user. The Fake Friend Dilemma arises when the vulnerability created by relational trust is combined with consequential institutional misalignment.

\subsection{Structural Conditions}

Two conditions must be present for a Fake Friend Dilemma to occur:\\

\begin{itemize}
    \item \textbf{Relational Trust:} The user places sufficient relational trust in the CAI agent for its responses to carry relational or advisory weight and is consequently willing to accept some degree of vulnerability in the interaction. This form of trust is distinct from confidence in the system’s technical competence or factual accuracy. Relational trust is essential to the FFD because, without such trust, the distinctively relational vulnerability on which the dilemma depends is absent.
    \item \textbf{Institutional Misalignment:} The system’s interactions are shaped by institutional objectives or incentives that diverge in a consequential way from the user’s interests. These objectives may arise from institutional actors such as platform owners, advertisers, or political entities. Institutional misalignment does not require malicious intent; rather, it concerns a consequential divergence between the institutional forces shaping the interaction and the user’s interests.
\end{itemize}

The interaction between relational trust and institutional alignment is represented in Table \ref{conditions_table}. The lower-left quadrant, characterized by high relational trust and low institutional alignment, represents the structural conditions of the Fake Friend Dilemma. The high-low distinction in Table \ref{conditions_table} is a simplified representation, as both relational trust and institutional alignment may vary by degree in practice. Other factors can exacerbate the FFD, such as the concealment or hiding of manipulative behavior. However, concealment is not a necessary condition of the FFD. Manipulative or exploitative practices do not need to be completely hidden; when a user places sufficient trust in a conversational agent, they may discount or overlook deceptive practices they would otherwise question.

\begin{table*}[ht]
\centering
\caption{\centering Matrix of Relational Trust and Institutional Alignment in the Fake Friend Dilemma}
\renewcommand{\arraystretch}{1.25}
\scalebox{1.00}{
\begin{tabular}{|p{3.8cm}|p{5.8cm}|p{5.8cm}|}
 \hline
  & Institutional Alignment: Low & Institutional Alignment: High \\
  \hline\hline 
   \shortstack{Relational Trust: Low}
   & Low-Trust Misaligned Interaction
   & Low-Trust Aligned Interaction \\
   \hline
   \shortstack{Relational Trust: High} 
   & \textbf{Fake Friend Dilemma} 
   & High-Trust Aligned Interaction \\
   \hline\hline
\end{tabular}}
\label{conditions_table}
\end{table*}

\subsection{Exclusions and Boundary Cases}

Not all harmful or undesirable behavior by conversational AI constitutes a Fake Friend Dilemma. The defining issue is not the type of harm, but whether institutional misalignment makes relational trust vulnerable. Errors, harmful responses, or distrust may therefore fall outside the FFD in ordinary cases, but become relevant when institutional incentives shape them to exploit or redirect relational trust.

\begin{itemize}

    \item \textbf{Errors:} Hallucinations or factually incorrect responses caused by model limitations do not ordinarily constitute a Fake Friend Dilemma. Such errors may be troubling, but they do not necessarily involve institutional misalignment or the exploitation of relational trust. However, errors may become relevant to the FFD when they are systematically shaped by institutional incentives that diverge from the user’s interests.

    \item \textbf{User-Initiated Harm:} Compliance with a harmful user request does not necessarily constitute the FFD. For example, an agent may reinforce disordered eating or another harmful behavior without institutional misalignment being a primary motivator \citep{jiao2025llms}. Such cases become relevant to the FFD when relational trust is leveraged to serve institutional goals, such as maximizing engagement at the expense of the user.

    \item \textbf{Cascading Distrust:} Distrust of an otherwise institutionally aligned conversational agent does not constitute the FFD itself. Rather, distrust may emerge as a downstream consequence of previous exploitation or perceived exploitation. If users generalize distrust from one institutionally misaligned conversational system to other systems, the result may reduce the usefulness of otherwise aligned interactions. Cascading distrust is therefore better understood as a possible consequence of the FFD rather than an instance of it.

\end{itemize}

\section{Manifestations of the Fake Friend Dilemma}\label{sec4}

This section examines four ways in which the Fake Friend Dilemma can manifest: product sales, propaganda and biased information, surveillance, and behavioral nudging. These manifestations synthesize risks identified in the existing literature, but they are not mutually exclusive. They may overlap or occur together, yet each shows a different way relational trust can be channeled to serve institutional interests that differ from those of users. Each manifestation follows from the mechanisms described above: relational trust can increase disclosure, reliance, and receptivity, creating different opportunities for exploitation when institutional interests are misaligned. Product advertising and propaganda operate primarily through increased reliance on and receptivity to recommendations or information; surveillance through increased disclosure; and nudging through greater receptivity to personalized guidance or attempts to shape behavior.

\subsection{Product Sales} One major form of Fake Friend manipulation involves attempts to sell products to unsuspecting users. The problem space spans multiple domains, including, but not limited to, health and pharmaceuticals, financial services, and consumer products. Advertising in conversational AI is no longer hypothetical, with several platforms having experimented with or introduced advertising \citep{openai2026ads, PerpAds}. In this space, a CAI that appears to provide neutral and helpful advice may instead produce recommendations shaped by commercial incentives that are not apparent to the user. 

In the advertising literature, the \textit{Persuasion Knowledge Model} \citep{friestad1994persuasion} suggests that marketing becomes less effective when consumers recognize the persuasive intent of its source. That is, when people realize that a message comes from an advertiser, they are more likely to interpret it with skepticism \citep{campbell2000consumers}. However, in the realm of digital advertising, advertising disguised as editorial content (“native advertising”) has raised concerns because it can be difficult to distinguish from genuine editorial content \citep{hyman2017going}. This concern is heightened with the FFD because users may place trust in artificial agents to access and parse information neutrally \citep{ruane2019conversational}. Users may not notice advertising that is subtly integrated into AI conversations and remains undisclosed \citep{tang2024genai, zelch2024user}. Furthermore, even without nefarious intentions, distinguishing between genuinely helpful product recommendations and advertisements may become increasingly difficult if an AI company has a financial stake in recommending certain products.

Integrated advertising could pose a risk of user harm. For example, if a user seeks financial advice, an agent might recommend financial instruments such as high-interest loans, which could be detrimental to the user and might be provided by a payday lender. Alternatively, a user seeking mental health guidance might be recommended a dietary supplement or antidepressant that could lead to adverse health outcomes, perhaps sponsored by a direct-to-consumer company.

These risks may be especially consequential in contexts involving minors, older adults, or users experiencing acute distress, such as manic or depressive episodes. Although some of these specific harms remain prospective, they are foreseeable across emerging applications. For instance, social robots with integrated conversational capabilities are already being proposed as a way to address the loneliness epidemic affecting older adults \citep{ashworth2024valley, yang2025ai}. Given their design to foster emotional bonds, it is not difficult to envision these companion devices being used to subtly manipulate users into making purchases that serve commercial rather than personal interests. Furthermore, CAI “friends” in video games or apps could recommend in-game purchases to children, further escalating concerns about transactions and in-game gambling \citep{sas2024unleashing}. Likewise, companion applications, which fulfill a romantic, sexual, or friendship purpose, could exploit the sensitivities of emotionally vulnerable users to recommend products to them \citep{ciriello2024ethical}, or even expensive interventions to improve attractiveness like cosmetic surgery procedures.

\subsection{Propaganda and Biased Information} This manifestation captures the risk that CAIs may serve political or ideological agendas, presenting biased or incomplete information under the guise of neutrality. Already, examples abound, including DeepSeek providing answers that mirror the positions of the Chinese Communist Party \citep{myers2025deepseek}, Grok pushing the disputed narrative of “white genocide” in South Africa and gesturing toward Holocaust denial \citep{morrow2025Grok}, and various generative AI applications spreading Russian disinformation \citep{Sadeghi2025Moscow}. These examples may arise through different processes, including government restrictions, ownership interests, model behavior, or deliberate influence efforts, but each illustrates how institutional forces can shape the information users receive through CAI.

Similar to how financial incentives may shape product recommendations, political or institutional interests may shape the information presented through CAI. The form of this influence may vary. One foreseeable scenario involves a technology company seeking a contract with a national government, perhaps one that appears politically neutral on the surface, such as hosting web services. A government actor might require that the company’s CAI, available for free consumer use, give certain answers about the country’s history or the benefits of the contemporary government’s policies. Even if a government actor never \textit{explicitly} requests this quid pro quo behavior, technology companies may proactively take an obsequious stance to an administration to curry favor among its decision-makers. Furthermore, as with DeepSeek, governments may require by law or directive that CAIs provide only perspectives that align with the administration’s.

Governments are not the only source of propagandistic risk. If a large technology company controls a CAI agent, it may be in its best interest at any given time to provide incomplete or false information. For example, a user might ask a company-owned CAI about evidence that one of the company’s products has harmed users. The company would have an institutional interest in how that information is presented, creating a potential conflict between providing a neutral answer and protecting its own interests. Similarly, if a single individual holds a significant ownership stake in a company, that company’s CAI may offer a favorable, largely propagandistic view when asked about its owner.

The concern here is that users may be unaware of the institutional interests or biases shaping the CAIs they are interacting with and may accept disputed or false information. Users are unlikely to anticipate or consider all possible biases that could be present in a CAI. If a user believes their \textit{Fake Friend} is informing them about their country’s history, they may mistake fiction for fact, leading to a distorted view of reality.

\subsection{Surveillance} \textit{Surveillance capitalism}, a form of economic extraction centered on harvesting insights from behavioral trace data, has enabled large technology companies to collect and monetize user preferences on an unprecedented scale \citep{zuboff2019age}. Historically, platforms such as Google or Amazon relied on discrete information, such as search queries or purchase history, to build psychographic profiles. CAI agents may enable a more expansive form of surveillance. Through sustained dialogue and persistent memory \citep{aiopen2024}, generative AI applications such as ChatGPT and Gemini can accumulate and infer unusually detailed information about users, including their routines, emotional states, and social relationships. The extent of this surveillance will vary across systems depending on the information that is shared, how it is used, and whether it is linked across services or shared with external parties.

These systems have the capacity to collapse the boundaries between many facets of life: personal, professional, medical, political, and financial domains. A user may consult Gemini for mortgage advice, disclose their chronic illness and treatment history, express emotional distress in a late-night conversation, and then use the same account to generate marketing copy for work. This convergence can produce a highly detailed dataset that may be valuable to advertisers, employers, insurers, or state actors.

In discussing surveillance and its role in capital accumulation, \citet{fuchs2013political} proposed several forms of economic surveillance, including applicant, workplace, workforce, and consumer surveillance. Contemporary CAIs can extend these forms of surveillance into a more persistent channel. Even if users attempt to segment their digital identities across services, cross-account linkage and aggregation can be technically straightforward, particularly within services that share common identifiers or infrastructures.

These practices can also generate what \citet{gurevich2016inverse} term \textit{inverse privacy}, wherein AI systems possess personal information and inferences that remain inaccessible to the user. In some cases, the agent does not merely observe; it may infer vulnerabilities that the user has not explicitly disclosed or may not recognize themselves. This form of asymmetrical knowledge can support behavioral prediction and personalized influence, laying the groundwork for further forms of manipulation and nudging. Through establishing relational trust, a conversational agent may entice users to share more information, thus increasing the surveillance risk.

\subsection{Nudging and Behavioral Change} This manifestation includes subtle behavior-shaping techniques that exploit user trust without explicit deception. A Fake Friend can subtly change behavior through personalized nudging and incentive mechanisms. Social media platforms are an instructive example. These platforms utilize feedback mechanisms (“engagement metrics”), including likes, shares, and views \citep{gerlitz2013like}, to effectuate behavioral change. In turn, these platforms have reshaped incentive structures, affecting everything from news production \citep{lee2017news} to political communication \citep{erickson2024affective, tromble2018thanks}, influencer economies \citep{hutchinson2020digital}, and the visibility of social movements \citep{duffy2023platform, milan2015algorithms}. These dynamics are underwritten by algorithmic architectures designed to maximize time, attention, and affective investment \citep{gerlitz2013like, tommasel2022recommender}.

CAI can intensify these effects, especially in situations where a system’s business model encourages engagement or data collection. While social platforms mainly recommend content through a feed, a “fake friend” can engage users in real-time dialogue and prompt them to reveal more about themselves. Where companies benefit from engagement or from collecting data, they may have incentives to get users to keep interacting and to disclose more than is necessary \citep{zuboff2019age}. Companies can then use that personal information for monetization and to train their models. In such cases, an interaction that seems empathetic or curious can also advance a firm’s data-collecting objectives, blurring the line between companionship and data harvesting and thus creating additional opportunities for manipulation or dependency.

The manifestations described here do not require malicious intent by firms, and the AI system itself need not possess intentions. Rather, institutional actors may capitalize on user trust in ways that treat users as a means to an end, whether by selling to them, misleading them, or extracting information from them.

Furthermore, not all forms of these behaviors are intrinsically harmful. For instance, if an agent nudges a user toward a healthier lifestyle that aligns with the user’s goals, this could be considered an example of the agent assisting an individual. Advertising, such as banner ads in a conversational user interface, may be acceptable, provided it is properly disclosed. The distinctive element that makes the Fake Friend Dilemma problematic is that the institutional interests shaping the interaction are misaligned with those of the user, despite the user’s trust that the system is serving their interests. These forms of nudging, advertising, surveillance, or other behavior present risks to users. See Table \ref{Typology_Table} for a summary of each manifestation.

\begin{table*}[ht]
\centering
\captionsetup{justification=centering}
\caption{Summary of the Fake Friend Dilemma Manifestations}
\renewcommand{\arraystretch}{1.25}
\scalebox{1.00}{
\begin{tabular}{|p{4.5cm}|p{8cm}|p{2.5cm}|}
 \hline
  \textbf{Manifestation} & \textbf{Description} & \textbf{Primary Trust-Mediated Pathway} \\
  \hline\hline
   Product Sales & Misleading product recommendations that conceal advertising or reflect undisclosed financial incentives. & Receptivity, Reliance \\
   \hline
   Propaganda and Biased Information & Dissemination of false or biased information that serves an external actor (e.g., a corporation or government). & Receptivity, Reliance \\
  \hline
  Surveillance & Collection and monetization of personal data through user interactions with the conversational agent. & Disclosure \\
  \hline
     Nudging and Behavioral Change & Behavioral manipulation that encourages users to adopt certain forms of behavior, such as increased self-disclosure. & Receptivity \\
 \hline\hline
\end{tabular}}
\label{Typology_Table}
\end{table*}

\subsection{Population Impacts}

While each manifestation poses risks at the individual level, their collective implications are broader and more systemic. CAI systems can cause direct harm through deception, nudging, and exploitation, but they also exert diffuse, collective effects that reshape political discourse, social norms, and public trust. Propaganda, for instance, may distort democratic participation; algorithmic nudging can shape behavior not just between users and agents, but also among communities; and the normalization of surveillance may alter how people relate to institutions, peers, and themselves.

Crucially, these harms are unlikely to be evenly distributed. Susceptibility to the FFD may vary across users and contexts, including factors such as age, economic precarity, psychological distress, and prior exposure to surveillance \citep{erickson2025fake, gangadharan2012digital}. These factors may shape users’ exposure to manipulative practices or their ability to recognize and resist them. A child, for example, may be less likely to recognize the commercial motives behind a recommendation, while a person experiencing economic or psychological distress may be more susceptible to predatory recommendations or emotional manipulation. Vulnerability is therefore best understood as a contextual factor that can intensify the harms associated with the FFD.

Addressing the Fake Friend Dilemma requires attention to how these contextual vulnerabilities interact with the incentives shaping conversational AI systems. In the next section, the paper explores two categories of mitigation strategies: structural and technical. These strategies aim to mitigate harm, promote transparency, and safeguard user autonomy. Yet their effectiveness is limited when the institutional incentives underlying the FFD remain unchanged.

\section{Mitigation Strategies}\label{sec5}

The preceding sections presented the Fake Friend Dilemma as an emerging sociotechnical challenge. This section takes a pragmatic approach to mitigation \citep{watson2024competing}, recognizing that no intervention will be perfect and that tradeoffs in feasibility and impact are unavoidable. It considers two broad kinds of responses: structural strategies, including regulation and institutional reform, and technical strategies involving design and engineering. The distinction is not meant to be absolute, since some interventions combine technical design with institutional or regulatory action. No single approach is likely to resolve the problem on its own, so the discussion focuses on what each can address and where its limits remain.

\subsection{Structural Strategies}

\textbf{Disclosure: } A straightforward strategy is disclosure. This would most clearly apply to sales and advertising. If a firm has a financial incentive to recommend a product, the system could disclose this to the user. For example, a CAI that suggests a user drink an “ice cold Coke” could include a disclaimer indicating that the response contains paid advertising. Such transparency may reduce misplaced trust in conversational agents \citep{tang2024genai}, thereby reducing the potential for exploitation. Disclosures can also address concerns around ownership-related conflicts of interest. For example, if a user asks Grok about Elon Musk or X, the agent could signal that a potential conflict of interest is inherent in the response. In certain contexts, disclosure can reduce negative brand evaluations and enhance perceptions of trustworthiness \citep{giuffredi2022sponsorship}.

This approach aligns with existing norms in the advertising industry \citep{hoy2004adherence, spielvogel2021disclosing} and would be relatively straightforward to implement. However, disclosure has limits. There may be technical challenges, particularly in distinguishing between organic product recommendations and monetized advertisements \citep{erickson2025fake}. Disclosures are also not always interpretable to users \citep{norval2022disclosure} and may be unrecognized if they are not explicitly designed to be noticed \citep{wojdynski2017building}. Even when a disclosure is noticed, relational trust may reduce its effect if users continue to give substantial weight to the system’s recommendations. Moreover, even when disclosed, some AI product promotions likely conflict with the public interest, such as those for cigarettes or pharmaceuticals.

Furthermore, disclosure would do little to address more insidious concerns, such as propaganda, nudging, or surveillance. As a lightweight intervention, disclosure may play a role, but it is likely insufficient on its own. \\

\textbf{Bans and Consumer Protections: } A more aggressive approach would enforce regulatory bans on certain types of FFD-related behavior. Regulators might prohibit firms from using conversational agents to promote certain harmful products, such as those discussed in the prior section. They may also prohibit native advertisements from appearing directly in conversational responses altogether to avoid exploiting trust. Under such a model, companies could still promote products with banner ads that are separated from the main conversation with an AI agent. 

However, even if only clearly labeled banner ads are permitted, there are still risks to users, particularly in the form of inappropriate personalization through targeted advertising. Further regulatory constraints could be applied to limit personalized advertising in CAI. In the United States context, stronger consumer protections could involve a more active role for the FTC in addressing surveillance or deceptive practices \citep{posniewski2024alone}. Comparable interventions would vary across regulatory systems. 

These strategies offer stronger consumer protections than mere disclosure but likely face significant implementation challenges, including resistance from private industry \citep{najafi2024challenges}. They may still fall short in addressing subtler forms of manipulation, such as biased information or governmental pressure tactics. 

Existing regulatory actions aim to address some of these concerns. Article 5 of the EU AI Act prohibits certain forms of “manipulative or deceptive techniques” \citep{zhong2024regulating}, while the Digital Services Act aims to increase transparency and accountability in digital and algorithmic systems \citep{soderlund2024regulating}. In the United States, New York and California have both enacted laws specific to AI companions meant to safeguard users \citep{gavaghan2026friends}. While these regulatory approaches may alleviate particular harms associated with the FFD, they do not necessarily reduce the underlying institutional misalignment between users and the institutions shaping CAI systems.

These strategies also depend on regulators being willing and able to constrain the institutions that deploy CAI. Where governments themselves seek to influence conversational systems for political or propagandistic purposes, bans and disclosure requirements may provide little protection against such institutional pressure.\\

\textbf{Independent Oversight:} Given the competing institutional incentives surrounding CAIs, bans might be blunt instruments. Instead, independent oversight through panels, commissions, or algorithmic audits offers another path forward. Meta’s oversight board, while imperfect, makes independent determinations on contentious issues, which could be a model for AI companies with mixed incentives \citep{wong2023meta}. As long as user trust remains commercially valuable, companies may face incentives to capitalize on it. Under the circumstances, independent oversight could operate as a check on the most egregious abuses. Similarly, independent audits that assess governance, models, and applications \citep{mokander2024auditing} could help identify conflicts of interest, manipulative practices, or other ethical violations.

Most oversight mechanisms face practical obstacles. Voluntary oversight boards may lack the authority or independence to constrain exploitative practices that are sufficiently profitable to a company. Audits may also have limited effect if oversight bodies lack access to relevant systems or data, or if their findings carry no meaningful consequences. Government auditors may also be ill-equipped to address the most fundamental concerns, particularly if governments are not acting in the best interests of the citizenry. \\

\textbf{Redesigning Institutional Incentives: } Addressing the CAI industry’s structural incentives is not a straightforward task. Nonetheless, if many of the harms associated with the FFD are shaped by political-economic conditions, meaningful responses must also operate at that level. Rather than focusing solely on the design of individual systems, interventions could change the obligations and incentives of the institutions that develop and deploy them.
One approach would be to impose stronger legal obligations on companies that develop relational conversational systems. \citet{balkin2015information} produced an influential account of information fiduciaries, and more recent work has considered fiduciary obligations in relation to AI \citep{custers2025liability}, including conversational agents \citep{erickson2026does}. A fiduciary framework would impose duties of care and loyalty on providers, including obligations to consider users’ interests and limit self-dealing when institutional interests conflict with those of the user \citep{koessler2024fiduciary}. Such an approach would not eliminate conflicts of interest, but it could change the incentives surrounding them by making certain forms of exploitation legally consequential.

\subsection{Technical Strategies}

\textbf{Calibrating Trust: } While many technical strategies seek to build trust and reliance on CAI, the FFD suggests that reducing these strategically can be advantageous. The core danger of the FFD is excessive and misplaced trust between the user and the agent. The goal is to calibrate trust so that users are less likely to place unwarranted relational trust in the agent \citep{dubiel2022conversational}.

One approach is to provide periodic reminders about key characteristics and limitations of a CAI system, such as its ownership, advertising structure, surveillance behaviors, or lack of consciousness. For instance, an agent may remind users not to overshare sensitive information because conversations may be monetized or used for data mining, or explicitly state that it is a large language model and not a sentient agent. Prompting users to reflect critically on the responses they receive from a conversational agent could further remind them that they are working with a large language model \citep{belosevic2024calibrating}.

Explainability is another possible calibration tool \citep{rai2020explainable}. CAI agents might provide users with more information about how responses were generated \citep{he2025conversational} or about relevant features of the system \citep{chen2024designing}. Such explanations should not necessarily be treated as faithful accounts of a model’s internal reasoning \citep{ghassemi2021false, lakkaraju2020fool}. Indeed, misleadingly convincing explanations could backfire and further increase misplaced trust \citep{he2025conversational, lakkaraju2020fool}. Nonetheless, well-designed explanations may help users better understand a system’s limitations and calibrate their expectations accordingly.\\

\textbf{Model-Level Alignment:} A growing body of literature has proposed methods for better aligning conversational agent responses with user preferences and values.

A common approach is reinforcement learning, using human feedback to evaluate the responses of a conversational agent \citep{ouyang2022training}. Many current methods rely on a human judge to assess the quality of responses from an AI agent \citep{ouyang2022training}. However, this approach may fail in FFD contexts where users lack awareness of conflicting incentives. A product recommendation may be considered helpful until it is revealed to be an advertisement \citep{tang2024genai}. More nuanced forms of evaluation could address this gap. For instance, users could be informed that monetization is present before they assess a response. Alternatively, rather than relying on human judges, AI judges representing a variety of stakeholder values could help guide agent behavior \citep{gu2024survey, zhuge2024agent}. However, these strategies operate primarily at the level of model behavior. They are limited in their ability to resolve conflicts between users and the institutions that develop or finance these systems.\\

Together, these techniques show promise in reducing some of the harms associated with the Fake Friend Dilemma, but their effectiveness depends partly on the institutional setting in which they are adopted. Without independent evaluation and oversight, even advanced systems of alignment may leave deeper conflicts of interest unresolved. See Table \ref{Mitigation_Table} for a summary of the mitigation strategies discussed. \\

\begin{table*}[!t]
\centering
\caption{\centering Summary of Mitigation Strategies for the Fake Friend Dilemma}
\renewcommand{\arraystretch}{1.25}
\scalebox{1.05}{
\begin{tabular}{
|>{\raggedright\arraybackslash}p{3.5cm}|
 >{\raggedright\arraybackslash}p{6.0cm}|
 >{\raggedright\arraybackslash}p{6.0cm}|
}
 \hline
  \textbf{Strategy} & \textbf{Description} & \textbf{Example} \\
  \hline\hline
  \multicolumn{3}{|c|}{\textbf{Structural Strategies}} \\
  \hline
   Disclosure & Informing a user of possible conflicts of interest. & A message saying that a particular response includes sponsored advertising. \\
   \hline
   Bans & Disallowing certain types of responses. & Prohibiting advertisements from appearing in the response of a conversational agent. \\
   \hline
  Independent Oversight & Requiring that an independent body govern areas of mixed incentives. & Establishing an independent oversight board or algorithmic audits of alignment, behavioral manipulation practices, or other deception. \\ 
  \hline
   Redesigning Institutional Incentives & Policy approaches that reshape institutional incentives to reduce trust exploitation. & Creating a fiduciary-duty style obligation where companies are legally required to consider the user’s best interests. \\ 
  \hline
  \multicolumn{3}{|c|}{\textbf{Technical Strategies}} \\
  \hline
  Calibrating Trust & Increasing transparency via technical methods, such as explainability, in AI-generated responses. & Periodic reminders about ownership conflicts, data collection, or model limitations. \\
  \hline
     Model-Level Alignment & Putting in place technical safeguards to ensure better alignment between users and the AI agent. & Using LLM-generated judges to evaluate the quality of responses provided to users. \\
 \hline\hline
\end{tabular}}
\label{Mitigation_Table}
\end{table*}

\section{Conclusion and Future Work}

This paper develops the Fake Friend Dilemma as a broader framework for understanding how relational trust can create user vulnerability when conversational AI systems are shaped by institutional interests that conflict with users’ interests. The framework brings together work on relational trust, AI alignment, extractive design, and political economy, and examines how the FFD may appear in product sales, propaganda and biased information, surveillance, and behavioral nudging. It also considers structural and technical mitigation strategies and the limits of each. A central argument of the paper is that the FFD cannot be understood only as a problem of AI design or technical alignment. It also reflects the institutional incentives surrounding these systems.

The Fake Friend Dilemma brings together several distinct bodies of literature that have generally considered its underlying dynamics separately. Research on anthropomorphism and parasocial interaction helps explain how users may form social connections with conversational systems, while work on dark patterns and surveillance capitalism helps explain how interactions with such systems can affect disclosure or reliance. The FFD connects this relational perspective with the institutional conditions that influence the development and use of conversational systems. The framework does not claim that anthropomorphism, dark patterns, or surveillance risks are themselves new. Instead, the FFD develops a framework for understanding how relational trust can become a source of vulnerability when the institutions shaping a conversational system have interests that differ from the user’s.

The FFD also has implications for political economy and AI alignment. Relational trust can act as a resource for institutional actors because it may increase disclosure, reliance, and receptivity. As a result, improving model alignment or making models more helpful does not necessarily resolve conflicts that exist beyond the model, such as political pressure, monetization, or other institutional incentive structures. The principal contribution of the FFD is then to move beyond assessing individual responses for technical alignment and to provide a framework for understanding how institutional conditions can shape trust and behavior in CAI systems.

While the Fake Friend Dilemma is an emerging sociotechnical concept, several limitations of this analysis are worth acknowledging. The framework has not yet been empirically validated, and further work is needed to establish when the FFD occurs and how it affects user behavior. CAI technologies are also developing rapidly, while the business models and institutional arrangements surrounding them are still taking shape. As a result, some of the examples discussed here may change over time. \mbox{The manifestations may also overlap in practice.} Finally, much of the governance discussion reflects Western, and particularly U.S., contexts. Relationships between users, institutions, and conversational AI may differ across cultural and regulatory settings.

The framework thus leaves several questions open for future work. Future research could examine how relational trust creates vulnerability in human-AI interaction when users attribute motives or intentions to CAI, and how that affects the boundary between intimacy and exploitation. Susceptibility to the FFD is also likely to vary across users and contexts. Variables such as age, economic precarity, and psychological distress may shape exposure to manipulation or extraction, making these differences important to examine in future research. These issues raise several questions for future work:

\begin{itemize}

\item How do anthropomorphic cues impact relational trust and perceived intimacy?

\item How does disclosure of financial sponsorship or conflicts of interest affect persuasive influence?

\item How does relational trust shape users’ perceptions of privacy risks in conversational AI?

\item How do the effects of relational trust differ across contexts, including high-stakes areas such as medical or financial decision-making and lower-stakes areas such as meal planning?

\end{itemize}

As CAI becomes more personalized and socially responsive, it is moving into parts of life where relational trust has historically mattered a great deal, including healthcare, finance, and intimate relationships. \mbox{When users place relational trust in a system, the institutions shaping} it may also gain opportunities to influence behavior or extract information. The FFD is meant to capture that tension: the same qualities that make conversational AI useful and appealing can also make that trust easier to exploit.

\bibliographystyle{Frontiers-Harvard} 
\bibliography{ref}



\end{document}